\documentclass{article}
\usepackage{arydshln}
\usepackage[hidelinks]{hyperref}
\usepackage{spconf,amsmath,graphicx}
\usepackage[table]{xcolor}
\usepackage{graphicx}
\usepackage{amsmath}
\usepackage{amssymb}
\usepackage{enumitem}
\setlist{nosep, leftmargin=14pt}
\usepackage{subcaption}
\usepackage{booktabs}
\usepackage{lipsum}
\usepackage{mwe} 

\title{CTI-Unet: Cascaded Threshold Integration for Improved U-Net Segmentation of Pathology Images}

\name{
    \begin{tabular}{ccccc}
        Mingyang Zhu$^{\star}$ & Yuqiu Liang$^{\dagger}$ & Jiacheng Wang$^{\ddagger}$\\
    \end{tabular}
}

\address{
    $^{\star}$ Shanghai Jiao Tong University, Shanghai, China \\
    $^{\dagger}$ Fudan University, Shanghai, China \\
    $^{\ddagger}$ Vanderbilt University, Nashville, USA
}

\begin{document}
%
\maketitle
\begin{abstract}

Chronic kidney disease (CKD) is a growing global health concern, necessitating accurate and efficient diagnostic tools. Automated segmentation of kidney pathology images is critical for facilitating precise diagnoses and treatment planning. However, conventional segmentation methods often struggle with the subjective selection of threshold values, where weak thresholds introduce noise and strong thresholds miss important details. In this paper, we propose a novel cascaded fine-tuning segmentation network that integrates outputs from multiple threshold levels to improve segmentation accuracy. Leveraging a U-Net-based architecture, our method effectively balances sensitivity and specificity by mitigating the limitations associated with single-threshold segmentation. Experimental results on the challenging KPIs2024 kidney pathology dataset validate the effectiveness of our approach, securing superior performance over state-of-the-art models such as nnU-Net, Swin-Unet, and Mamba-Unet. Our code and models will be made publicly available to facilitate further research.
\end{abstract}


\section{Introduction}
Chronic kidney disease (CKD) is a progressive condition characterized by the irreversible decline of kidney function, often resulting from diabetes, hypertension, and other diseases that alter renal structure over time~\cite{Webster2017Chronic}. Between 2000 and 2019, kidney diseases ascended from the 19th to the 10th leading cause of death worldwide, with mortality increasing significantly during this period~\cite{WHO2021Top10}. CKD now affects more than 10\% of the global adult population, and this prevalence continues to rise~\cite{Ammirati2020Chronic}. Individuals of lower socioeconomic status are disproportionately affected, experiencing a higher risk of CKD progression compared to those of higher status~\cite{Ammirati2020Chronic}. The associated complications—including anemia, fatigue, comorbidities, and depression—severely diminish patients' quality of life~\cite{Morton2023Quality}.

Accurate diagnosis and effective treatment planning for CKD rely heavily on histological examination of kidney tissue~\cite{Hogan2016Native}. Kidney biopsy remains the gold standard for diagnosing CKD, providing critical insights into the structural alterations within the kidney. However, the increasing demand for biopsy interpretation is met with a shortage of qualified pathologists, highlighting the need for automated tools to assist in the analysis of kidney pathology images.

In this work, we utilized Periodic acid-Schiff (PAS) stained whole slide images (WSIs) from the Kidney Pathology Image Segmentation (KPI) challenge to develop a novel convolutional neural network model based on U-Net, aiming to achieve precise pixel-level segmentation of glomeruli. Automated tissue segmentation plays a pivotal role in the quantitative analysis of kidney biopsy specimens. Precise segmentation enables the identification and quantification of pathological features, facilitating better understanding and management of the disease.

Building on recent advancements, our work introduces a novel algorithm based on the U-Net architecture. This method leverages the strengths of U-Net while addressing specific challenges observed in previous studies, such as the subjective selection of threshold values in segmentation tasks.

\section{Related Work}
Recent advancements in artificial intelligence (AI), particularly in deep learning, have introduced powerful tools for image segmentation tasks. Convolutional neural networks (CNNs) have been widely adopted in medical image analysis. The U-Net architecture~\cite{Ronneberger2015UNet} is one of the most prominent CNN models designed for biomedical segmentation, providing pixel-level classifications and enabling precise localization of anatomical structures. Since its introduction, numerous variants of U-Net have been proposed to improve segmentation performance. nnU-Net~\cite{isensee2021nnu} is an automated framework that adapts U-Net configurations to given datasets, achieving state-of-the-art results across various medical imaging tasks. Swin-Unet~\cite{cao2022swin} incorporates Swin Transformers to capture long-range dependencies, while CE-Net~\cite{gu2019net} introduces context encoding and spatial attention mechanisms.

Despite the success of these models, challenges remain in achieving optimal segmentation performance on complex pathology images. A significant issue is the subjective selection of threshold values for binary segmentation. Weak thresholds may include excessive noise, while strong thresholds can lead to the loss of important details. This subjectivity introduces variability and inconsistency in segmentation results, which is undesirable in clinical applications.

To address these challenges, some approaches have explored multi-thresholding techniques~\cite{sezgin2004survey}, combining outputs from different thresholds to balance sensitivity and specificity. However, integrating multiple thresholds within a deep learning framework for medical image segmentation remains underexplored.

\begin{figure}[htbp] 
\centerline{\includegraphics[width=1\linewidth]{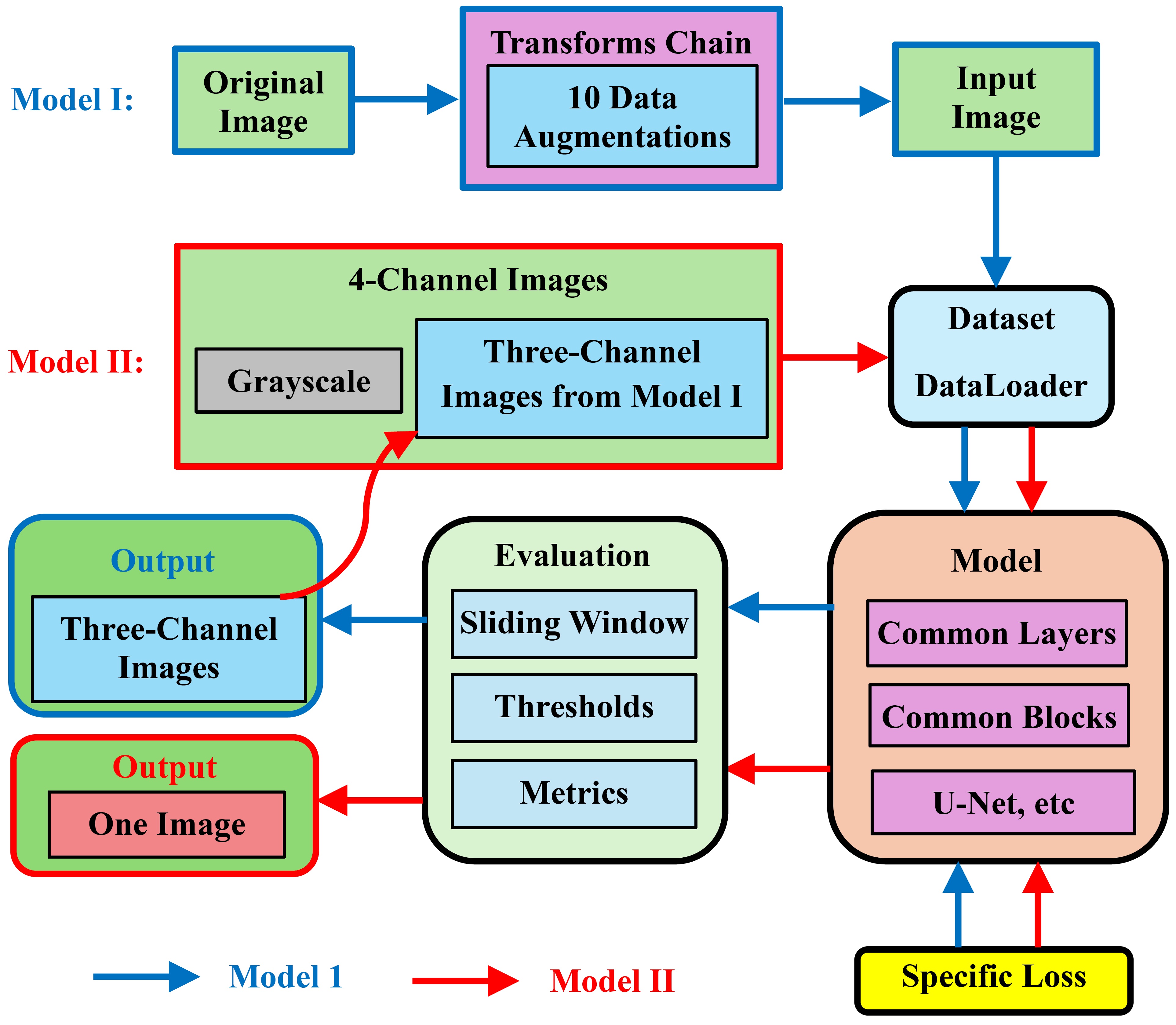}} \caption{Overview of the proposed Cascaded Threshold-Integrated Segmentation Framework. The initial segmentation module provides preliminary masks, which are processed at multiple thresholds. The Threshold Integration Network then refines these outputs to produce the final segmentation.} 
\label{fig:framework} 
\end{figure}

\section{Methods}
\subsection{Overview}

We propose a novel \textbf{Cascaded Threshold-Integrated U-Net (CTI-Unet)} that enhances segmentation accuracy by smartly integrating outputs from multiple threshold levels. Our method is built upon the standard U-Net architecture and specifically designed to address challenges in segmenting PAS-stained kidney images.

The CTI-Unet consists of two main components:

\begin{enumerate} 
\item \textbf{Initial Segmentation Network (Model 1)}: Generates preliminary segmentation masks using the U-Net architecture. 
\item \textbf{Threshold Integration Network (Model 2)}: Refines the initial segmentation by integrating outputs from multiple thresholds, effectively balancing sensitivity and specificity. 
\end{enumerate}

By decoupling the initial segmentation from the threshold integration, our framework offers flexibility and can be adapted to various segmentation tasks and models. The key innovation lies in the Threshold Integration Network, which leverages the strengths of different thresholds to overcome the limitations of single-threshold segmentation.

\subsection{Initial Segmentation Network (Model 1)}

The Initial Segmentation Network is responsible for generating preliminary segmentation masks from high-resolution kidney pathology images. We utilize the standard U-Net architecture~\cite{Ronneberger2015UNet}, implemented using the Medical Open Network for AI (MONAI) framework~\cite{cardoso2022monai}.

\begin{figure}[h] 
\centerline{\includegraphics[width=0.9\linewidth]{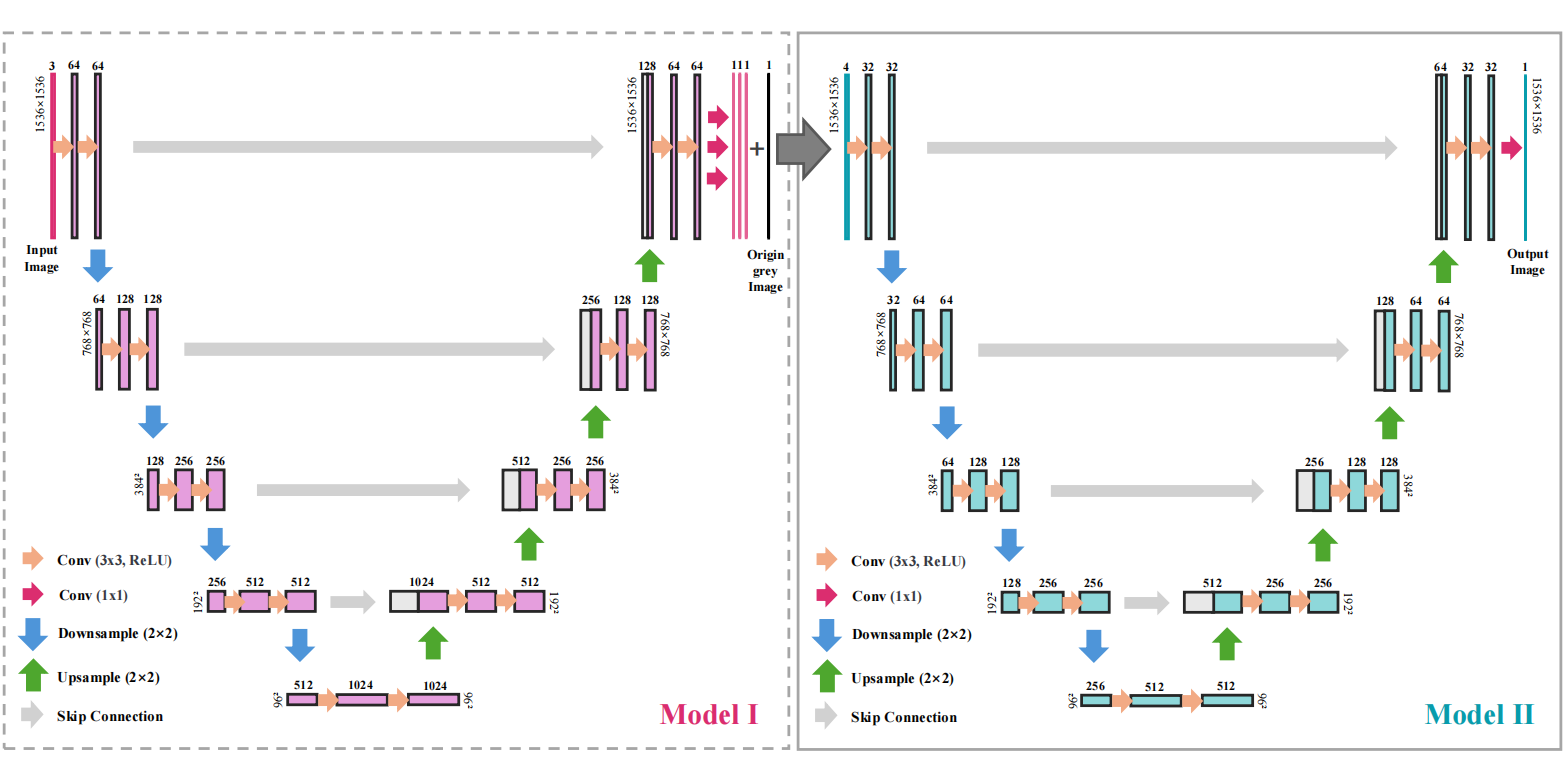}} \caption{Network architecture of our prosed CTI-Unet. Left pink part is the initial segmentaion network while the right cyan part is the threshold integration network.} 
\label{fig:network} 
\end{figure}

As shown in Fig.~\ref{fig:network} left part colored in pink, the network follows the typical encoder-decoder structure with skip connections to preserve spatial information. Input RGB images of size $2048 \times 2048$ pixels are processed to generate segmentation masks.

To enhance the model's generalization capability, we apply extensive data augmentation techniques during training, including random rotations, zooms, flips, affine transformations, and intensity adjustments (see Table~\ref{table:augmentation}).

\subsection{Threshold Integration Network}

The Threshold Integration Network refines the initial segmentation by integrating outputs from multiple thresholds. This network takes as input the original image and the set of preliminary segmentation masks obtained at different thresholds (e.g., 0.01, 0.1, 0.6).

As illustrated in Fig.~\ref{fig:network} right part colored in cyan, Model 2 is also based on the U-Net architecture but modified to accept multiple input channels. The input consists of a grayscale version of the original image and the three segmentation outputs from Model 1 corresponding to the different thresholds, resulting in a 4-channel input.

The network learns to balance sensitivity and specificity by analyzing the relationships among the different threshold outputs, effectively reducing noise and recovering details.

\subsection{Implementation Details}
\textbf{Data Preprocessing}: Input images were normalized to have zero mean and unit variance. We resized images to $2048 \times 2048$ pixels and converted them to grayscale for Model 2.

\noindent\textbf{Data Augmentation}: We applied data augmentation using MONAI transforms (see Table~\ref{table:augmentation}).

\noindent\textbf{Training Details}: Both networks were trained using the Adam optimizer with a learning rate of $1 \times 10^{-4}$. We employed a batch size of 4 and trained each model for 100 epochs.

The loss function is a combination of Tversky Loss and Cross-Entropy Loss where $\alpha_1 = 0.7, \beta_1=0.3$ and $\alpha_1 = 0.5, \beta_1=0.5$:

\begin{equation} \mathcal{L} = \mathcal{L}{\text{Tversky}}(\alpha_i, \beta_i) + 0.5 \times \mathcal{L}{\text{CE}}. \end{equation}

where equal weighting of $\alpha$ and $\beta$ emphasizes a balance between false positives and false negatives.

\noindent\textbf{Inference Procedure}: During inference, we apply a sliding window approach with overlap to handle large images. Model 1 generates segmentation outputs at multiple threshold levels, which are then fed into Model 2 for refinement.

\begin{table}[htbp]
\centering
\resizebox{\linewidth}{!}{%
\begin{tabular}{|l|l|c|}
\hline
\textbf{Transform} & \textbf{Parameters} & \textbf{p} \\ \hline
Random Rotate 90° & Axes: [0, 1] & 0.5 \\ \hline
Random Zoom & Min zoom: 0.9, Max zoom: 1.1 & 0.5 \\ \hline
Random Axis Flip & Flip along random axes & 0.5 \\ \hline
Random Affine & Rotate value: 0.1, Scale value: 0.1, Shear value: 0.1 & 0.5 \\ \hline
Random Grid Distortion & Distort limit: 0.03, Num cells: 5 & 0.5 \\ \hline
Random Gaussian Noise & Mean: 0.0, Std: 0.1 & 0.2 \\ \hline
Random Adjust Contrast & Gamma range: (0.7, 1.5) & 0.5 \\ \hline
Random Shift Intensity & Offset range: (0.1, 0.2) & 0.5 \\ \hline
Random Histogram Shift & Num control points: 3 & 0.5 \\ \hline
Random Gaussian Smooth & Sigma range: (0.5, 1.0) for x, y, z & 0.5 \\ \hline
\end{tabular}%
}
\caption{Data augmentation during our training process. p indicates the probability of applying an augmentation.}
\label{table:augmentation}
\end{table}




\section{Experiments and Results}
\subsection{Dataset}

We evaluated our method on the KPIs2024 dataset \cite{tang2024holohisto}, which includes high-resolution PAS-stained WSIs of kidney tissue with various conditions, such as 5/6 nephrectomy (\textbf{5/6Nx}), diabetic nephropathy (\textbf{DN}), nephrotoxic serum nephritis (\textbf{NEP25}), and normal kidney samples. Each WSI is of size $2048 \times 2048$ pixels at 40x magnification.

\subsection{Experimental Setup}

Our experiments were conducted on a single NVIDIA RTX 3080 GPU with 16GB of VRAM. The dataset was split into 80\% for training and 20\% for validation.

\subsection{Evaluation Metrics}

We used the Dice Similarity Coefficient (DSC) and Intersection over Union (IoU) as the primary evaluation metrics to assess segmentation performance. These metrics measure the overlap between the predicted segmentation masks and the ground truth annotations.

\subsection{Results and Analysis}

We compared our proposed framework with three state-of-the-art segmentation models: nnU-Net~\cite{isensee2021nnu}, Swin-Unet~\cite{cao2022swin}, and CE-Net~\cite{gu2019net}. All models were trained and evaluated under the same conditions for a fair comparison.

Table~\ref{table:quantitative} summarizes the quantitative results across different kidney conditions.

\begin{table}[htbp]
\centering
\resizebox{\linewidth}{!}{%
\begin{tabular}{|l|c|c|c|c|c|}
\hline
\textbf{Method} & \textbf{5/6Nx} & \textbf{DN} & \textbf{NEP25} & \textbf{Normal} & \textbf{All} \\ \hline
nnU-Net & 90.50 & 84.25 & 88.40 & 92.00 & 88.79 \\ \hline
Swin-Unet & 91.10 & 85.50 & 89.00 & 93.00 & 89.65 \\ \hline
CE-Net & 91.75 & 86.10 & 89.80 & 93.50 & 90.29 \\ \hline
Our Method & \textbf{93.14} & \textbf{87.79} & \textbf{91.41} & \textbf{94.73} & \textbf{91.64} \\ \hline
\end{tabular}%
}
\caption{Performance comparison of our method with nnU-Net, Swin-Unet, and CE-Net on the KPIs2024 dataset. DSC (\%) scores are reported for each kidney condition.}
\label{table:quantitative}
\end{table}

As shown in Table~\ref{table:quantitative}, our method achieves the highest DSC across all kidney conditions, with an overall DSC of 91.64\%, outperforming the compared models. The improvements are particularly notable in challenging conditions such as NEP25 and DN.

\subsection{Qualitative Results}
\begin{figure}[htbp]
    \centering
    \begin{subfigure}[b]{0.20\textwidth}
        \centering
        \includegraphics[width=\textwidth]{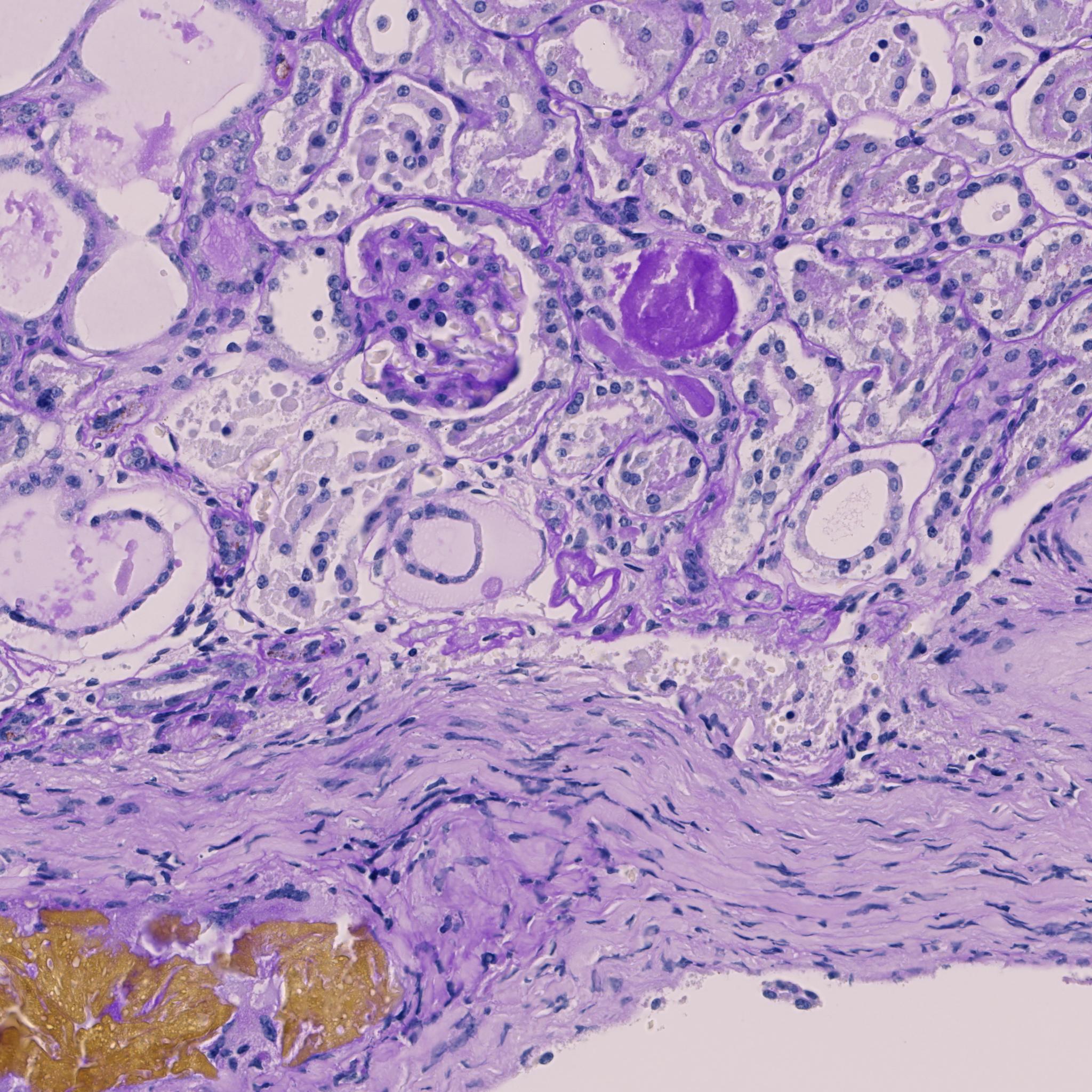}
        \caption{Original Image}
        \label{fig:original}
    \end{subfigure}
    \hfill
    \begin{subfigure}[b]{0.20\textwidth}
        \centering
        \includegraphics[width=\textwidth]{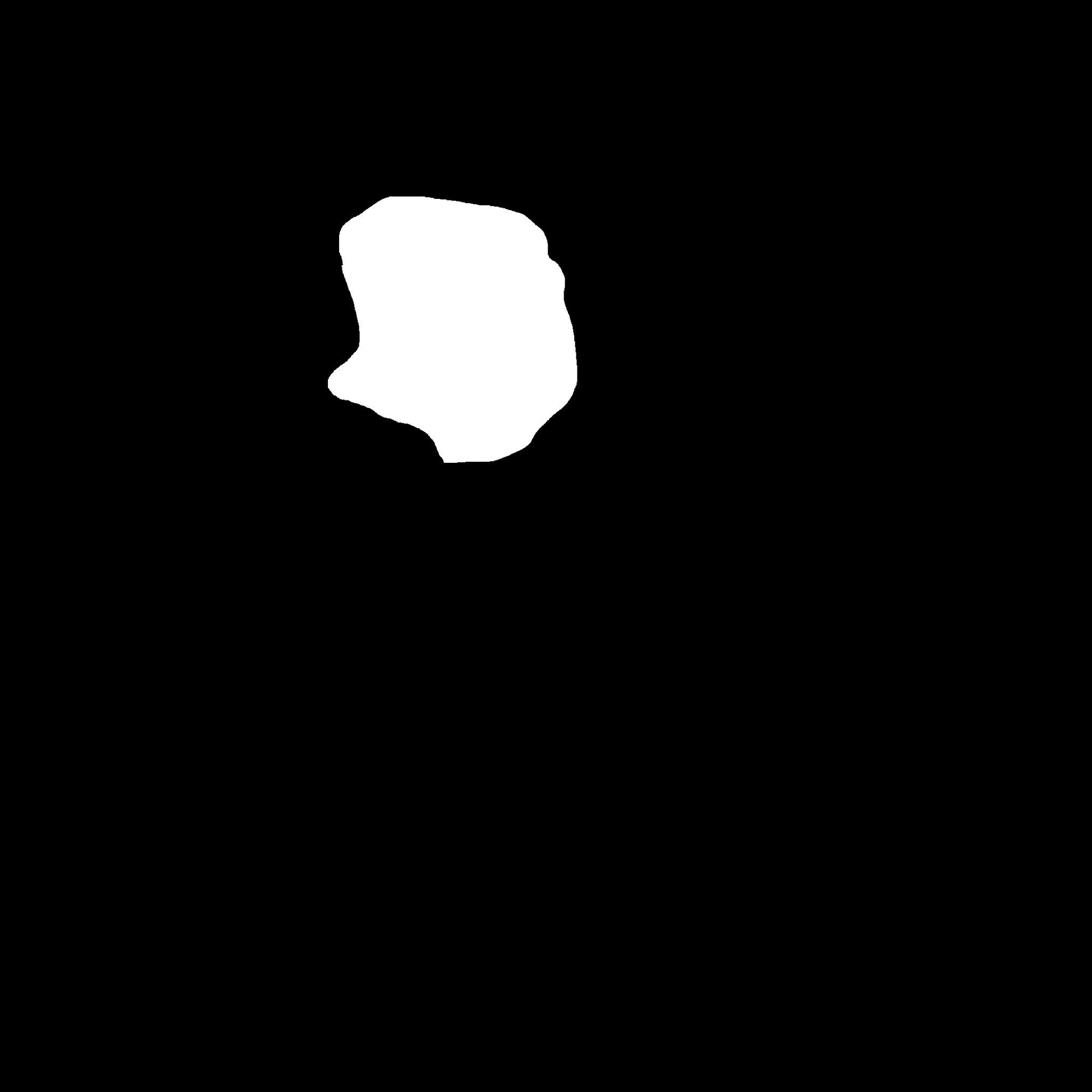}
        \caption{Ground Truth}
        \label{fig:true}
    \end{subfigure}
    
    \vspace{0.3cm}
    \begin{subfigure}[b]{0.20\textwidth}
        \centering
        \includegraphics[width=\textwidth]{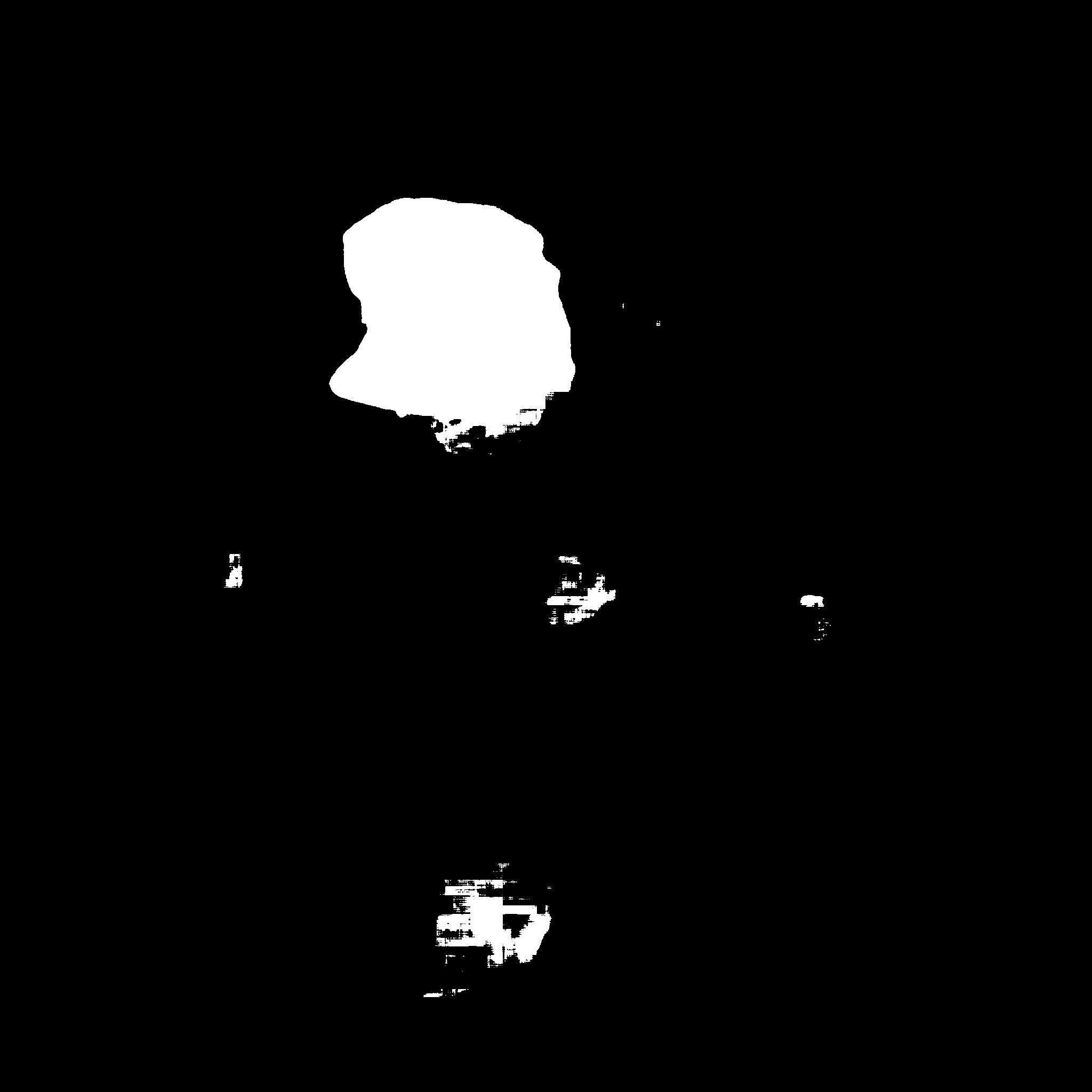}
        \caption{Model 1 Output}
        \label{fig:model1_output}
    \end{subfigure}
    \hfill
    \begin{subfigure}[b]{0.20\textwidth}
        \centering
        \includegraphics[width=\textwidth]{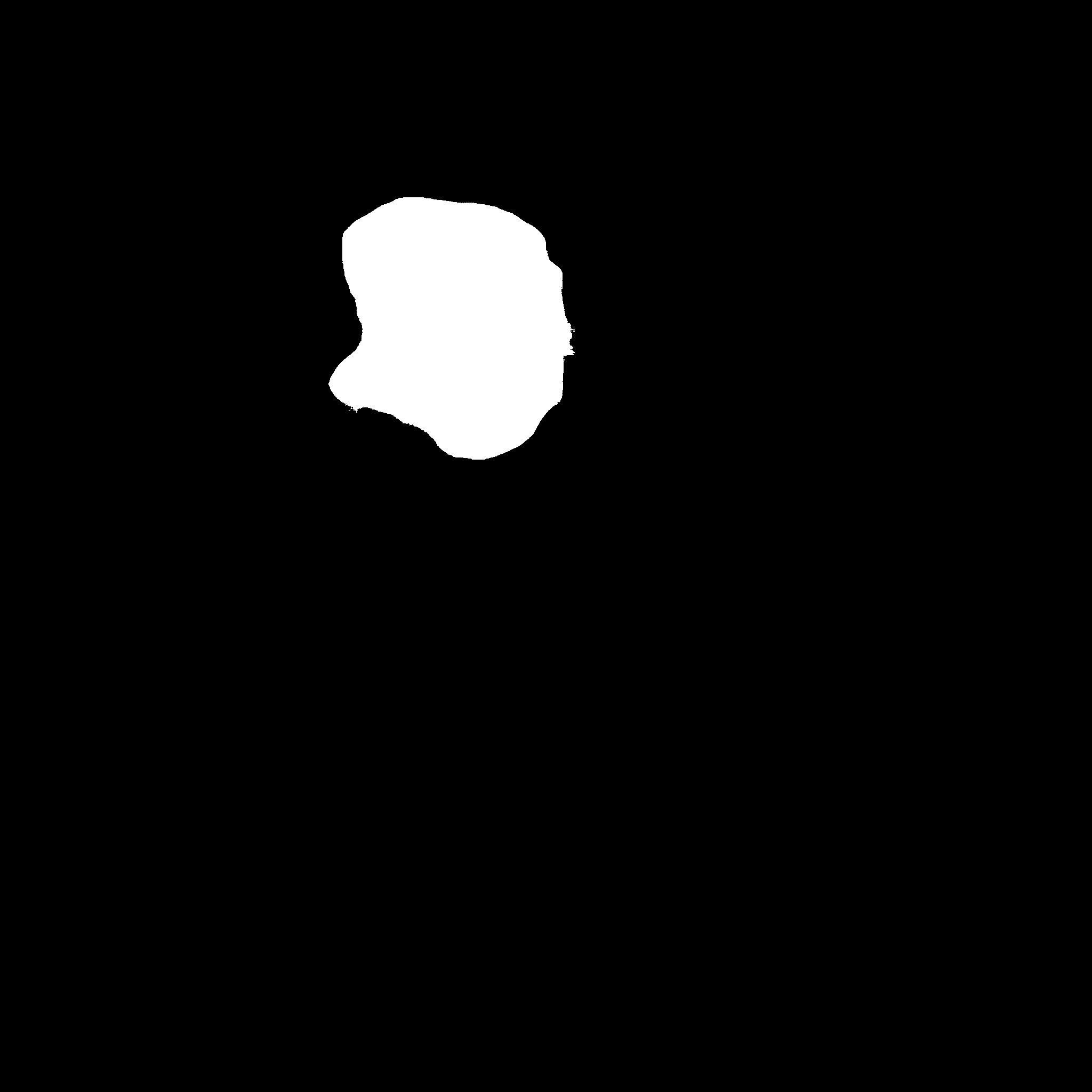}
        \caption{Model 2 Output}
        \label{fig:model2_output}
    \end{subfigure}
    
    \caption{Comparison between the original image, ground truth, and the outputs of Model 1 and Model 2. As shown, Model 2 improves upon Model 1 by reducing noise and generating smoother boundaries, which can be seen by comparing the lower two images.}
    \label{fig:comparison}
\end{figure}

The qualitative results Figure~\ref{fig:comparison} show that Model 2 effectively reduces noise and captures finer details, leading to segmentation outputs that closely match the ground truth.

\section{Conclusion}
We have introduced a novel \textbf{Cascaded Threshold-Integrated U-Net (CTI-Unet)} that enhances the segmentation of PAS-stained kidney pathology images by smartly combining outputs from multiple threshold levels. Our method addresses the challenge of subjective threshold selection in segmentation tasks, providing a more robust and accurate tool for analyzing complex pathology images. Experimental results on the KPI challenge dataset demonstrate that our approach outperforms the baseline U-Net model.

Future work will explore the integration of advanced attention mechanisms and transformer-based architectures within the Threshold Integration Network to further enhance feature representation and segmentation performance. Additionally, validating our method on other challenging medical imaging datasets will further establish its robustness and generalizability.

\noindent\textbf{Compliance with Ethical Standards.}
This research study was conducted retrospectively using human subject data made available in open access by MICCAI 2024 KPI Challenge. Ethical approval was not required as confirmed by the license attached with the open access data. 

{
\bibliographystyle{IEEEbib}
\bibliography{ref.bib}
}

\end{document}